# Effects of the COVID-19 Pandemic on Population Mobility under Mild Policies: Causal Evidence from Sweden

Matz Dahlberg[1,2,3,4,8], Per-Anders Edin[2,3,4,8], Erik Grönqvist[2,3,8], Johan Lyhagen[5,8], John Östh[4,7,8,9]*‡, Alexey Siretskiy[6,8], Marina Toger[4,7,8]

**Affiliations**:

1. Institute for Housing and Urban Research, Uppsala University, Uppsala, Sweden.

2. Department of Economics, Uppsala University, Uppsala, Sweden.

3. Institute for Evaluation of Labour Market and Education Policy, Uppsala, Sweden

4. Urban Lab at Uppsala University, Uppsala, Sweden.

5. Department of Statistics, Uppsala University, Uppsala, Sweden.

6. NevoLogic AB, Uppsala, Sweden

7. Department of Social and Economic Geography, Uppsala University, Uppsala, Sweden.

8. CALISTA centre for applied spatial analysis, Uppsala University, Sweden.

9. Jheronimus Academy of Data Science, ´s-Hertogenbosch, The Netherlands

‡ Correspondence to: E-mail: john.osth@kultgeog.uu.se

* The authors appear in alphabetical order.

**Abstract:**

Sweden has adopted far less restrictive social distancing policies than most countries following the COVID-19 pandemic (1–7). This paper uses data on all mobile phone users, from one major Swedish mobile phone network, to examine the impact of the Coronavirus outbreak under the Swedish mild recommendations and restrictions regime on individual mobility and if changes in geographical mobility vary over different socio-economic strata. Having access to data for January-March in both 2019 and 2020 enables the estimation of causal effects of the COVID-19 outbreak by adopting a Difference-in-Differences research design. The paper reaches four main conclusions: (i) The daytime population in residential areas increased significantly (64 percent average increase); (ii) The daytime presence in industrial and commercial areas decreased significantly (33 percent average decrease); (iii) The distance individuals move from their homes during a day was substantially reduced (38 percent decrease in the maximum distance moved and 36 percent increase in share of individuals who move less than one kilometer from home); (iv) Similar reductions in mobility were found for residents in areas with different socioeconomic and demographic characteristics. These results show that mild government policies can compel people to adopt social distancing behavior.

**Introduction**

On January 31, 2020, the first case of the new Coronavirus (COVID-19) was detected in Sweden (a person that had visited the Wuhan area in China) (8). On February 26, the second COVID-19 case was detected in Sweden (a person that had visited northern Italy) (9). From February 27 and onwards, more cases were detected, but the growth rate in new cases was initially fairly low.[1] These dates corresponds well with the first peaks observed in Google trends in Sweden between January 1 and April 2, 2020 for the search string "coronavirus" (see the red line in Figure 1), thus indicating that a general awareness of the new virus in the Swedish population started by the end of January (which seems to be the case for most countries in the world).

**Figure 1**: Google trends on search string "coronavirus", in Sweden and internationally

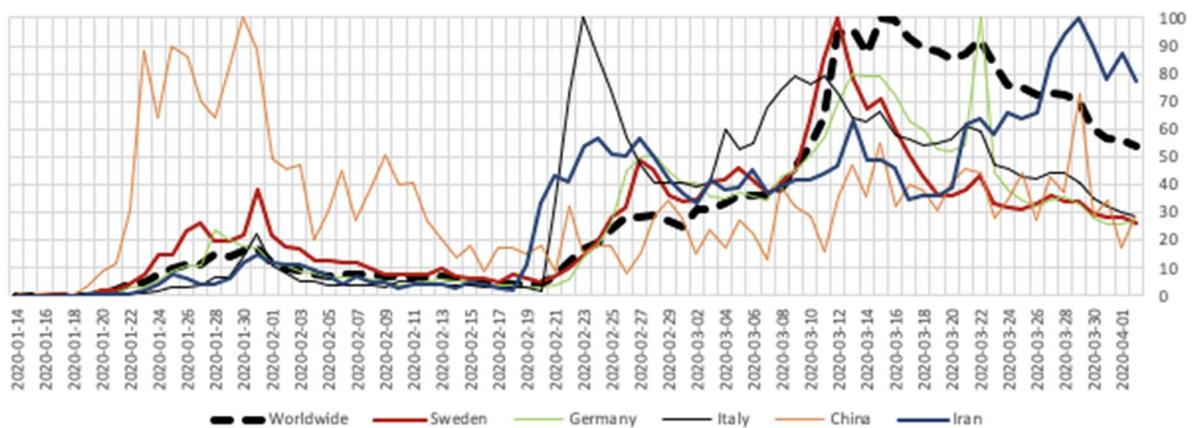

Source: Google trends.

Sweden is exceptional compared to most other countries in the world, in its quite different response to the COVID-19 outbreak by relying on individuals acting responsibly and adhering to public recommendations. This outlier position has led to international attention and debate (1–7). While several other countries have adopted quite drastic policies to stop infectivity at a larger scale, Sweden has adopted a route of less restrictive social distancing policies and mild recommendations following the Coronavirus epidemic.[2] By March 26, The Public Health Agency of Sweden's (PHAS)

---

[1] See supplementary table 1 in the Supplementary material for a timeline of the development of the Coronavirus in Sweden.
[2] By March 26, the date used for the analysis in this paper, Sweden's closest neighbours and the major economies of Europe had imposed social distancing by adopting policies that implied a full, or close to full

recommendations and policy suggestions[3] were still relatively mild: (i) public gatherings of more than 500 individuals not allowed, (ii) all citizens with a cold or influenza-like symptoms asked to remain at home, (iii) individuals older than 70 years instructed to limit contact with other persons as far as possible, (iv) employers with the possibility to let their employees work from home recommended to consider recommending that, (v) all education in high schools and at Universities changed to distance teaching, (vi) non-essential travels within Sweden recommended to be avoided, (vii) individuals without symptoms instructed to continue with their usual daily activities (even if family members are infected by COVID-19).

There are two features of the Swedish system put forward as contributing to why Sweden has responded very differently from other countries. First, experts and expert authorities, rather than politicians, have an important say in the decision-making process. Sweden has a long tradition of technocratic policy-making, where expert authorities and expert commissions have an important role to play. As Steinmo (2013) puts it: "there are few countries in the world that have relied as heavily on expert commissions to address complicated or politically difficult policy issues" (17). Second, Swedish citizens show a large (and increasing) trust in the political and administrative system (17–19).

Maintaining a social distance is key to reducing the rate of transmission of the virus (20,21). But can social distancing be reached by voluntary measures in a society with a high degree of social capital? To assess the effectiveness of the policy, aggregate mobility data is vital to understand how aggregate flows of people have changed (22), but even if such data provides unique opportunities[4], there are also important methodological challenges that need to be addressed: e.g. sample selection problems in cases when individuals share the data voluntarily, participation and attrition related to the policy measures of interest, or time-variant patterns in mobility. This study analyzes the effect of

---

lockdown (March 9-11 for Italy (10,11); March 11 for Denmark (12); March 12 for Norway (13); March 16 for Spain and France (12); March 18 for Finland (14); March 22 for Germany (15); and March 25 for the UK (16)); see Supplementary table S2 in the supplementary material for a more detailed description of the policies adopted in different countries.

[3] The Public Health Agency of Sweden is the expert Agency in Sweden that has the leading role in informing the general public on the COVID-19 development and in providing recommendations and suggesting public policies to combat the spread of the virus. Table S1 provides a timeline showing the Agency's views and standpoints on the Coronavirus at different points in time. It can be noted that March 26 is well beyond both the dates when the more draconian policies had been adopted in other countries in Europe and when Google trends peak for Sweden on March 12 (on March 11, the WHO declares the spread of COVID-19 a pandemic and on March 13 the PHAS declares that the COVID-19 situation in Sweden has entered a new phase with a high risk of community spread).

[4] Important work in Italy and China (23,24) use mobile phone data to document mobility patterns using mobile phone data.

mild policies, and is the first (to the best of our knowledge) to relate representative mobile phone users and their mobility patterns against a baseline in normal times and to adjust for seasonal patterns in mobility.

**Aim of the study**

The aim is to estimate the causal impact of the COVID-19 outbreak under the Swedish mild recommendations and restrictions regime on individuals' choice of geographic location and geographic mobility, and to assess if the observed change in geographical mobility varies over different socio-economic strata.

**Methods and data**

To provide a causal answer to the research question, a Differences-in-Differences (DiD) methodology is employed to estimate the behavioral responses to the spread of the COVID-19 and government policies between January 16, 2020, and March 26, 2020, using detailed mobile phone data in the greater Stockholm area. Individuals' mobility patterns in mid-January, before the first confirmed Swedish case (cf. Supplementary table S1) and when the general awareness of COVID-19 was still limited in the Swedish public (cf. Figure 1), is compared to the situation on March 26, with 2,840 confirmed COVID-19 cases and 77 fatalities (25) and when the Swedish government had instigated a number of relatively mild recommendations and restrictions (cf. Supplementary table S1). Any general difference in mobility patterns between mid-January and end of March, not related to the Corona epidemic, is accounted for by netting out the corresponding differences between January 17 and March 28 2019. Under the assumption that the time difference in 2019 represents the counterfactual change in mobility in 2020, the DiD estimand will here capture the causal impact of virus spread and government actions on behavioral responses (26,27). Throughout the analysis, we use Thursdays as the weekday of study (Thursday is the "most typical" day (28)).

Mobile phone data is used for the greater Stockholm area with a resident population of 2,594,759 people.[5] The underlying information, on which the analysis is based, is the location of all mobile phones every 5 minutes serviced by one of the major Swedish mobile operators. Each phone's locations can only be tracked for a maximum of 24 hours. For this analysis phone's locations were aggregated to hourly data by 1km×1km squares for the four dates. Sweden had a market penetration of mobile phones of 1.27 per person in 2018 (29), and 98 percent of Swedish households in 2017 had a mobile phone (30). As most individuals carry one phone, the number of observed phones is approximated with individuals in the analysis.

---

[5] The geographical space is limited by latitude 58.86-60.10 and longitude 17.20-19.45, and population counts are derived from the latest available population registers provided by Statistics Sweden (December 31, 2017).

Figure 2 describes the mobile phone density 10-11AM on January 16, 2020 (the left map shows the larger Stockholm area, including Uppsala, and the right map zooms in on the central parts of Stockholm).[6] In total, the location of more than half a million mobile phones are observed. The highest density, with more than 10,000 individuals per square kilometer, is observed in central Stockholm with a high concentration of commercial and business activities. High density levels are also observed in an octopus-like pattern along the main commuting routes (subway, commuter train and highways) out from the center, where a high mass of business and industrial areas are located. Residential areas are mainly located in the circle around Stockholm city center. The red arms out from the Stockholm city center mark commuting routes with urban conglomerations. The high density of individuals in the north-west corner is the university town of Uppsala (4th largest urban area in Sweden).

---

[6] The reason for using 10-11AM is to have a time point that is off the main commuting hours and the main lunch hours. January 16 is chosen as the "normal" pre-Coronavirus date; i.e. after the main Christmas holidays in Sweden and before individuals in Sweden were aware of the Coronavirus (c.f. Figure 1 and Table S1).

**Figure 2**: Maps of mobile phone densities, day population 10-11AM on January 16, 2020 (larger Stockholm area including Uppsala in left map, central parts of Stockholm in right map).

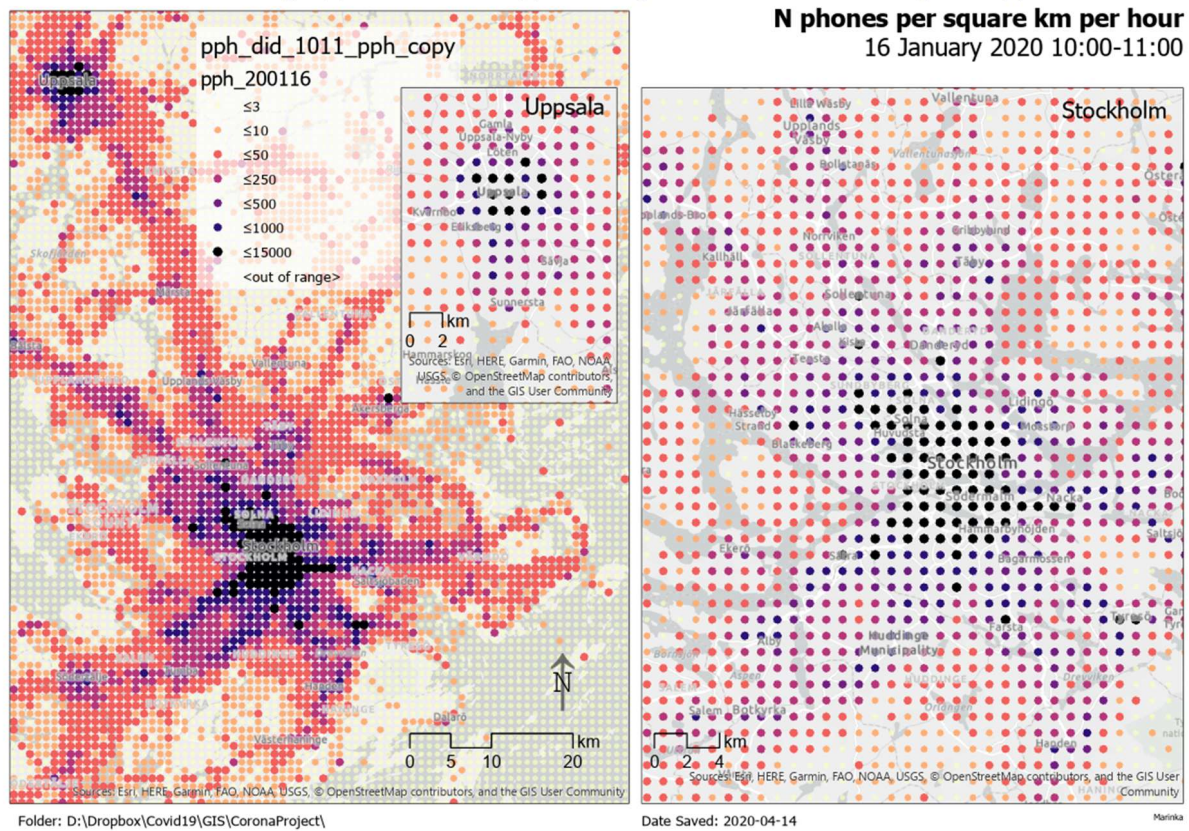

**Results**

Two types of results will be presented. First, changes in the geographic distribution of individuals (mobile phones) at a certain time point at a certain day. Second, changes in individual (mobile phone) mobility.

*Change in population densities*

The geographic distribution of mobile phones at 10-11AM on Thursday March 26, 2020, is compared with the corresponding distribution at 10-11AM on Thursday January 16, 2020. To account for any constant changes in mobile phone distributions between January and March, the March-January difference in 2020 is compared to the corresponding difference in 2019 (effectively adopting a DiD design). From the results, presented in Figure 3, a clear decrease in the central parts of Stockholm and Uppsala can be noted (c.f. left map in Figure 3), as well as along the main, octopus-like, commuting routes observed in Figure 3 (see, e.g., the blue diagonal line between Uppsala and Stockholm in Figure 3, corresponding to the main commuter stretch between these two cities). Zooming in on Stockholm (right map in Figure 3), a clear decrease in the central parts of Stockholm is visible while at the same time there is a strong increase in individual densities in the residential areas surrounding the central parts of Stockholm (c.f. the maps in Figure 2).

**Figure 3**: Map of changes (Difference-in-Differences design) in mobile phone densities. Differences constructed and displayed in the maps are: [(day population, 10-11AM, Thursday March 26, 2020) – (day population, 10-11AM, Thursday January 16, 2020)] – [(day population, 10-11AM, Thursday March 28, 2019) – (day population, 10-11AM, Thursday January 17, 2019)].

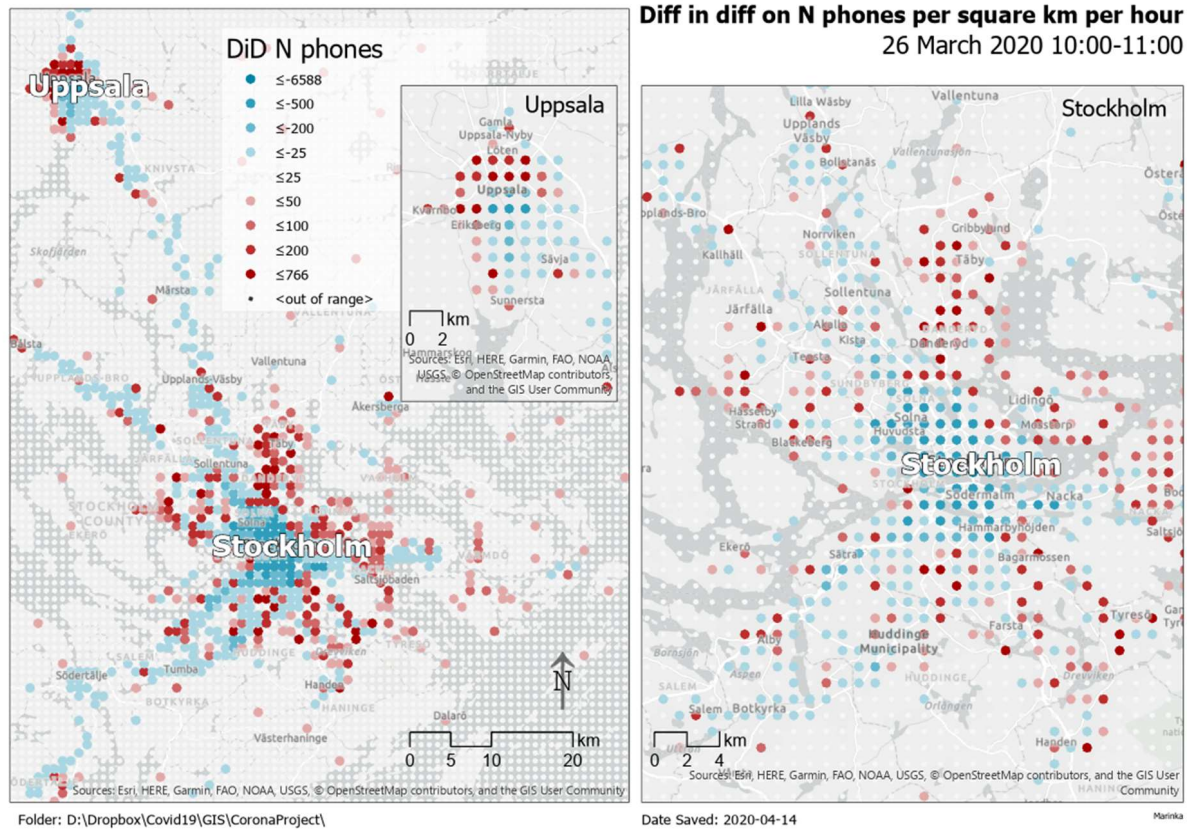

The maps of Uppsala and Stockholm in Figure 3 indicate an increase in individual densities in residential areas and a decrease in business and commercial areas. To get a more precise measure of what characterizes the areas with increasing and decreasing densities, respectively, we construct Moran's I local index of spatial autocorrelation (MI LISA) (31) to identify areas with statistically significant increases (so-called hot, HH, spots) and decreases (so-called cold, LL, spots) based on the DiD estimates provided in Figure 3 (see the map in Figure 4).[7]

---

[7] LISA MI identifies statistically significant clusters of low (cold spots) and high values (hot spots) using permutation bootstrap tests to determine how these clusters cannot reasonably be reproduced by chance. See supplementary material for more details on how the Moran's I LISA was constructed.

While the general pattern of lower densities on roads and in downtown Uppsala and downtown Stockholm is observable from the estimates in Figure 3, the MI LISA analysis clearly reveals the locations of the significant cold spots in the mixed employment and commercial central urban areas (the blue areas in Figure 4) and the significant hot spots in the low density suburban areas (the red areas in Figure 4).[8] This is also illustrated by comparing the distribution of distances to the 100 nearest jobs for the HH and LL area (the box-whisker plots in the right graph in Figure 4; in the HH areas individuals have to travel a longer distance to reach the 100 nearest jobs than the individuals in the LL-areas[9], indicating that they live in more residential areas.

**Figure 4**. Map of Moran's I local index of spatial autocorrelation, showing the areas with statistically significant increases and decreases based on the DiD estimates provided in Figure 3 (see graph on the left) and box-whisker plots showing the distribution of the distances to the 100 nearest jobs for the areas with significant increases and decreases, respectively.

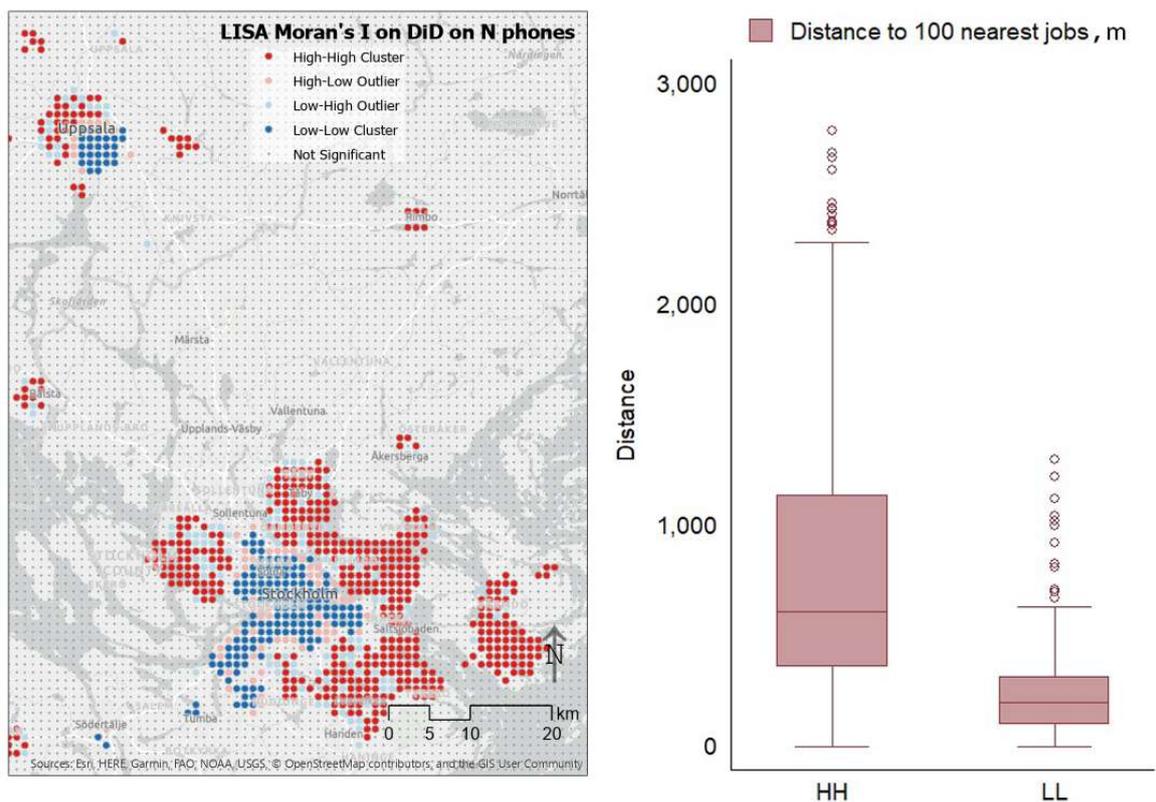

---

[8] Compare from supplementary figure 3 in the Supplementary material, which maps the area classification in the CORINE Land Cover database (32) for the area under study in this paper.
[9] The mean (median; max) value is 808 meters (608 meters; 2786 meters) in HH-areas compared to 244 meters (200 meters; 1300 meters) in LL-areas.

To get a sense of the magnitude of the changes, we calculate the average percentage changes of the DiD estimates in the HH and LL areas in the Moran's I map in Figure 4.[10] On average, there was a 64 percent increase in the number of phones in the HH areas and a 33 percent decrease in the LL areas.

*Changes in individual mobility*

To get a more direct measure of how individuals move, the maximum distance each mobile phone moves from its home during 24 hours is calculated. The home location was deduced from the night rest location of the mobile phone, and was calculated for 54 percent of the phones in the data (see supplementary material for definition). Table 1 shows that on January 16, 2020, the maximum distance travelled from home was on average 6,172 meters, and by March 26, 2020, the DiD estimate shows that this distance had been reduced with 2,346 meters: a 38 percent reduction. This change is mainly caused by a substantial increase in the number of immobile individuals.

Figure 5 depicts the distribution of the maximum distances moved from one's home on January 16 and March 26, 2020, respectively, and shows a clear shift of the distribution to the left by March 26 with a larger share of very small distances moved: the share of individuals who move less than one kilometer from home increases from 0.36 to 0.49 (a 36 percent increase) over the period.[11] At the same time the share of individuals moving 5-10 kilometers from home is reduced from 0.15 to 0.10 (a 33 percent decrease) and the share that travels 10 kilometers or more is reduced from 0.20 to 0.13 (a 35 percent decrease); cf. Supplementary table S3.

An interesting feature of the decreased mobility is that the effect seems homogenous across areas with different socioeconomic characteristics.[12] Table 1 shows that the decrease in mobility is the same as the overall effect in areas with the 10 percent highest concentration of residents belonging to visible minorities. The same is true for areas with the 10 percent highest concentration of residents having a university degree, living in relative poverty[13], or belonging to the risk age group (above 70).

---

[10] The DiD estimates are related to their pre-COVID19 outbreak values (the January 16, 2020, values).
[11] The large share of phone users moving less than one kilometer on January 16, 2020, can be related to the fact that 14.7 percent of the population are 75+, and that only 60.5 percent of the population above 15 are employed (33,34).
[12] Using adjacent registry data on socioeconomic characteristics, the residents in each geographical area (1km square) can be characterized and linked to the night rest location of the mobile phone.
[13] Poverty is defined as having a disposable income below 60 percent of the full population aged 16-74.

**Figure 5**. Distribution of meters travelled from home during the 24 hours on March 26 and January 16, respectively.

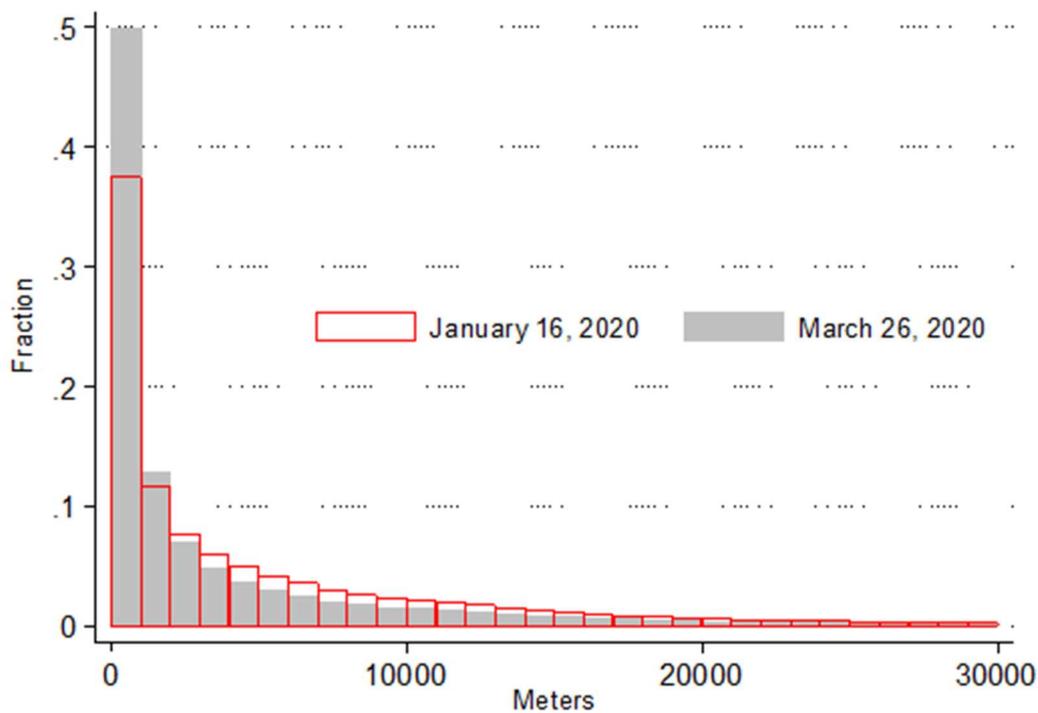

**Table 1**. Average of the maximum distance from home on January 16 2020 and the DiD-estimate of the change between January 16 and March 26, 2020: Overall and for geographical cells with a high density (90th percentile) visible minorities, high educated, low income and risk group.

|  | All | Minority (p90) | High Education (p90) | Poor (p90) | Risk Group (p90) |
|---|---|---|---|---|---|
| Maximum distance from home Jan 16 (meters) | 6172 | 6065 | 4690 | 6169 | 5930 |
| DiD estimate of change in max distance (meters) | -2346 | -2272 | -1917 | -2355 | -2260 |
| *Percentage change* | *-38.0* | *-37.5* | *-40.9* | *-38.2* | *-38.1* |
| Number of observations | 280,522 | 28,431 | 28,808 | 28,109 | 28,336 |

**Discussion**

Sweden is an outlier country in its reliance on relatively mild recommendations and restrictions in response to the Coronavirus. By exploiting a DiD strategy, the causal impact of the COVID-19 outbreak on aggregate mobility is estimated for the larger Stockholm area using representative mobile phone data from one of the major Swedish mobile operators. The analysis reaches four main conclusions. (i) The daytime population in residential areas increased significantly: a 64 percent average increase in the number of individuals in the areas with the largest increases in daytime population (Hot spots). (ii) The daytime presence in industrial and commercial areas decreased significantly: there was a 33 percent average decrease in the areas with the largest decreases in daytime population (Cold spots). (iii) The distance individuals move from their homes during a day was substantially reduced: there was a 38 percent decrease in the maximum distance moved and a 36 percent increase in the share of individuals who move less than one kilometer from home. (iv) Similar reductions in mobility were found for residents in areas with different socioeconomic and demographic characteristics (visible minorities, highly educated, poor, and being 70 years or older).

The results show that even mild public recommendations and restrictions can compel people to practice social distancing, such as staying away from crowded areas and avoiding unnecessary travel. For example, more people stay in close proximity to their home, complying with the recommendation by the PHAS to work from home if possible. Hence, adopting a more lenient strategy based on individuals' judgement and sense of responsibility can be an effective strategy in a society with a large social capital and trust in public expert authority. In addition, there can be benefits to the more lenient strategy as it allows for flexibility (e.g. allowing individuals to take care of important matters, and enabling parents to continue to work by keeping elementary schools open).

An interesting and puzzling finding is that the change in mobility does not differ across areas with different socio-economic characteristics (visible minorities, highly educated, poor, and being 70 years or older). This is at odds with the increasingly popular narrative in the media reporting a clear socioeconomic gradient in the social distancing response (35). It is also at odds with the fact that 6 of the 15 first fatalities in Stockholm belonged to the same minority group and resided in the same neighborhood (36), potentially suggesting that there are area-based heterogeneities in mobility response. A potential explanation is that this analysis measures the responses to the COVID-19 outbreak on March 26. Even if by that date the severity of the epidemic and the necessity to take precaution might have been conveyed to most societal groups, still, the speed of adopting the social distancing practices might have varied across various groups. There can also be other sources of heterogeneity (e.g. housing conditions). We consider this to be an important question for future analyses.

The analysis shows a non-negligible mobility response in Sweden to the COVID-19 contagion and social distancing policies relying (mainly) on voluntary actions. This is an important lesson for policy as it may not be necessary to close down the whole economy to achieve social distancing. But it is of course still an open question whether these responses are large enough to slow down the spread of the disease. Moreover, the anecdotal evidence, which must be interpreted carefully (e.g. evidence based on Google information which may suffer from lack of comparability due to selection and seasonality issues)[14], suggests that the social distancing is somewhat smaller in Stockholm than in similar cities with more stricter policies in neighboring countries.

---

[14] Google community report for March 29, 2020, suggests local presence in residential areas has increased with 7 percent in Stockholm as compared to 10 and 14 percent in Copenhagen and Oslo, whereas presence at transit stations decreased with 45 presence in Stockholm compared to 55 and 64 percent in Copenhagen and Oslo (37).

**Funding**: MT and JÖ acknowledge support from *Axel och Margaret Ax:son Johnsons Stiftelse för allmännyttiga ändamål*. **Author contributions**: M.D., P-A.E., E.G., J.Ö. designed research; M.D., P-A.E., E.G., J.L, M.T., J.Ö. performed research; M.D., P-A.E., E.G., J.L, M.T., A.S., J.Ö. analyzed data; M.D., P-A.E., E.G., J.L, M.T., A.S., J.Ö. wrote and edited the paper. **Competing interests**: None of the authors have relationships or activities that could appear to have influenced the submitted work.
**Data availability**: The data used in the analysis will be made available to any researcher for replication purposes.

# Supplementary materials (methods, tables and figures) for

# Effects of the COVID-19 Pandemic on Population Mobility under Mild Policies: Causal Evidence from Sweden

Matz Dahlberg[1,2,3,4,8], Per-Anders Edin[2,3,4,8], Erik Grönqvist[2,3,8], Johan Lyhagen[5,8], John Östh[4,7,8,9]*‡, Alexey Siretskiy[6,8], Marina Toger[4,7,8]

**Affiliations**:
1. Institute for Housing and Urban Research, Uppsala University, Uppsala, Sweden.
2. Department of Economics, Uppsala University, Uppsala, Sweden.
3. Institute for Evaluation of Labour Market and Education Policy, Uppsala, Sweden
4. Urban Lab at Uppsala University, Uppsala, Sweden.
5. Department of Statistics, Uppsala University, Uppsala, Sweden.
6. NevoLogic AB, Uppsala, Sweden
7. Department of Social and Economic Geography, Uppsala University, Uppsala, Sweden.
8. CALISTA centre for applied spatial analysis, Uppsala University, Sweden.
9. Jheronimus Academy of Data Science, ´s-Hertogenbosch, The Netherlands

‡ Correspondence to: E-mail: john.osth@kultgeog.uu.se

## Supplementary Methods

*Geographical setting and generation of geo-coded data*
Two main databases are used in this study, the MIND database and the PLACE database. In the subsequent text, the databases, the data drawn and the variables constructed will be described.

*Geographical variables calculated from MIND-data*
MIND is the name of the database that holds all phone data that is used for analysis.

When turned-on, each phone that is within reach of cell-tower-service will be connected to a cell-phone element located on a cell-tower. Cell-towers are more common in areas with greater concentrations of homes, services or jobs. This means that the geographical spread of towers varies substantially but is closely aligned to the population distribution.

The information from cell-towers is being preprocessed, involving operations making detailed tracking of each person impossible, and joined with the geographical position of cell-towers. For that, the join field was constructed based on Location Area Code (LAC), Service Area Code (SAC) for 3G and Tracking Area Code (TAC), Evolved Node B (ENB) and Sector ID for 4G-LTE. Thus, one can get an estimation about cell phones' mobility at the geographical resolution of the position of cell towers. The positional data granularity, i.e. the minimal spatial object, corresponds to the size of Service Area (SA) circa several hundred meters. Due to the large volume of data (roughly a half billion observations per day), the calculations were carried out in a parallel manner on a Spark Cluster operating on top of a distributed file system (HDFS). This allowed to reduce the preprocessing delivery time from about 20 hours to about 1-1.5 hours, and offering a straightforward way to further increase performance in the future by adding more nodes to the cluster.

*Variables describing mobility*
Data from the MIND database enables tracing phone locations as they switch from one cell to another. Since the time of any event is known (at five-minute granularity) and the location of each cell tower is known, several time-space relevant variables were constructed.

*Calculating mobility distances*
First step for locating a phone and calculating the distances moved by any user is to use the coordinate of the cell tower as the position of the phone user. As the phone user moves in space and the phone-service is switched from one tower to another tower, a motion can be registered and stored. However, instead of assuming that each phone moves between the towers (the service does but not the phone), we assume that the phone moves to the half-way distance intersection point between the cell-towers. In the subsequent steps of mobility calculations, new intersection points are determined by finding the half-distance between the last intersection point and the new cell tower servicing the user. The calculation is exact enough for the construction of usable mobility data, but also coarse enough not to identify detailed trajectories of the phone-user[1]. The procedure allows us to mimic mobility trajectories that come closer to factual mobility, and that is more well distributed geographically compared to distributions using cell tower locations alone. Using this technique, the maximum distance moved away from the assumed coordinates of origin (residential coordinates) during the course of the day is calculated.

---
[1] We may not trace any phone for more than 24h.

*Aggregate coordinates created for data merging*
The above estimation of the phone locations were aggregated to square km units and by hour. These aggregations were used as indexes for joining data from different datasets to each other, but it is central to understand that variables expressing the individuals' travel distances were preserved. For coordinates (x as an example) all spatial observations (on metric scale) were truncated using the following formula:

$$XKM = trunc(X/1000)*1000 + 500$$

This means that all events were *relocated* to the midpoint of the 1km x 1km grid. For aggregation timewise all events that take place within the same hour were classified into that hour (e.g. the events occurring at 09:35 and at 09:15 both got the hour value of 9). It should be noted that a phone user could move between the grid units within the same hour, therefore contributing to data in more than one square and hour. Variables that are explained in the subsequent text will make different use of these conditions.

*Coordinates of Origin*
Night rest position or Origin is calculated to determine the home location for each phone. We collected coordinates, using the method described above, for each individual phone between hours 03:00 AM and 06:55 AM, and weighted the coordinates by the duration of service (the time each phone remained under the service of a particular cell tower). The result is the OWX and OWY (Origin Weighted X and Y) coordinates for each phone, equivalent to the duration weighted average coordinate. The residential coordinates are truncated to km² units using the same formula as above.

The following Geographical variables created and used in the dataset are described in Table S4.

*Socio-economic variables calculated from population register PLACE*
*Data*
All population variables are created using the geo-coded population register drawn from the central bureau of statistics Sweden generated PLACE database. The register contains data for the full population with detailed geocoded information about income, education, age, and country of origin for each individual, and where the residential coordinates are known on a 100 m² accuracy. The database contains longitudinal population data but the last date of observation is December 31, 2017. Since the mobility study relies on phone mobility during winter and spring 2020 and corresponding dates in 2019, we make use of the latest 2017 material in the depiction and contextualisation of the phone population.

*Method*

In order to describe the population composition at the midpoint of each km² we are using a bespoke k-nearest neighbor technique. The choice of a bespoke neighbor technique rather than just using the population resident with each km² is motivated by a number of reasons. <u>First</u>, compared to k-nearest computations, fixed-area measures risk having several measurement problems related to the so called boolean border and MAUP (1–3). In Supplementary Figure S4, the first of the listed biases is illustrated in the two km² areas (grid areas A and B, containing 10 x 10, 100m² units each - equivalent to the geocoded level of detail in the population register). Let us assume that the different colours represent demographic qualities we want to associate with the cell-towers and km². The illustrated cell tower (X) is providing service to the encircled population, however by aggregating the population data on km² level (unit named A), a different set of demographic qualities are assigned to the area and the cell tower.

<u>Second</u>, the k-nearest neighbor approach ensures that the same count of neighbors is used for all computations and that the statistics are not based on too few individuals. In km²- areas B,C,D, the demographic ratios would be based on varying population counts, where the ratio in km² area C would be based on only two 100 m units.

The bespoke neighborhood approach makes uses a different technique for matching population data to the phone data. In the subsequent description the various steps will be presented:
1. Using the bespoke neighborhood tool EquiPop flow (https://equipop.kultgeog.uu.se/) we calculate the share of a selection of demographic qualities around each populated 100m2 unit. For each location the 100 nearest neighbors (full or adult population depending on variable) are searched and the demographic ratio is calculated for each populated coordinate. Using spatial join(in ARCGis pro), we match the rendered population computation values to the midpoint of the KM2 on the basis of proximity. The procedure ensures that the population variables represent the population composition at each KM2 midpoint.
2. Demographic qualities we are including in this part of the analyses include
    a. Belonging to an age-risk-group (full population). In Sweden all individuals >= 70 years of age, as well as all individuals having different health conditions are indicated as risk-groups that all other individuals should try to reduce contact with as much as possible. We have access to age as a variable (but not health conditions) so we calculate the share of individuals being aged 70 or older among the 100 and 500 nearest neighbors from each coordinate.
    b. Belonging to visible minorities (full population): Individuals born in Africa, Asia (minus Russia), and Latin America are listed as visible minorities. The underlying reasons for this classification are manifold and include: risk of discrimination on labour and housing market. In addition, recent years' migration increase is strongly associated with VM individuals, with poor language skills, residential segregation, etc., are more common in these immigrant groups compared to other groups. Under the current circumstances - poor integration, lower skills in Swedish, discrimination and segregation may lead to information deficits relating to recommended precautions and potential risks.

c. Belonging to the poor group (full population). Using the EU definition of relative poverty (9) we categorize the population having a disposable income lower or equal to 60% of the mean disposable income as poor.
   d. Belonging to a High-Education group (Adult population). Individuals having any tertiary education are listed as higher educated. The used values represent the share of higher educated adult individuals among the 100 nearest adult neighbors.

*Selection of high-high and low-low cluster areas using Moran's I LISA*

Moran's I local index of spatial autocorrelation (4) was constructed using ESRI's ArcGIS Pro software function Cluster and Outlier Analysis Anselin Local Moran's I using neighborhood definition of inverse distance weighting (IDW) with max distance threshold of 3 km. The 3 km distance threshold was selected so the resulting neighborhood for the 1 km gridded data corresponds to rook contiguity neighborhood of up to the third order. Smaller distances did not pick up the neighborhood effect on such a grid and larger thresholds gave similar results to the 3 km. The input variable was the Diff-in-Diff of day population phone densities as described in Figure S3 and Methods sections. The number of permutations was 499 but other higher numbers gave similar results.

## Supplementary tables

**Supplementary table S1**. The Public Health Agency of Sweden is the expert Agency in Sweden that has the leading role in informing the general public on the Coronavirus (COVID-19) development and in providing recommendations and suggesting public policies to combat the spread of the virus. This table provides a timeline showing the Agency's views and standpoints on the Coronavirus at different time points.

| Date | Information from the Public Health Agency of Sweden (PHAS) |
|---|---|
| Jan. 16 | The PHAS reported that a new Coronavirus had been detected among people that had visited a market in Wuhan in China. The Agency considers the risk that the new virus spread to Sweden to be very low |
| Jan. 30 | The WHO classifies the outbreak of the Coronavirus as a Public Health Emergency of International Concern (PHEIC). The PHAS states that a few cases in Sweden is not unlikely to be seen, but still considers the risk for infectivity in the Swedish society as low |
| Jan. 31 | First case of Coronavirus detected in Sweden (woman in Jönköping; had visited Wuhan area in China) |
| Feb. 5 | February 5: The PHAS still considers the risk for infectivity in the Swedish society as low |
| Feb. 13 | The PHAS still considers the risk for infectivity in the Swedish society as low. They also consider individuals' voluntary behavioral changes (e.g. in terms of international travels) to be preferred over state-imposed restrictions |
| Feb. 20 | The PHAS reports a large demand for information on the new Coronavirus. They provide information in Swedish and English and report that the Swedish information has had 800,000 page visits. The Agency still considers the risk for infectivity in the Swedish society as very low |
| Feb. 25 | The PHAS's new assessment of the Corona situation is that the risk for detecting COVID-19 cases in Sweden is high, but that the risk for general infectivity in the Swedish society is low |
| Feb. 26 | Second case of Coronavirus detected in Sweden (a person in Gothenburg that had visited northern Italy). The risk for general infectivity in the Swedish society is still considered as low |
| Feb. 27 | New cases of individuals infected with COVID-19 detected in Sweden |

| March 1 | The PHAS do not think that closing schools is an efficient policy. Based on existing outbreaks of Coronavirus in other countries, they find it unlikely that healthy kids cause infectivity. They also state that there exists no clear scientific evidence showing that closing schools/not allowing healthy kids to go to school would decrease the risk for infectivity in society even if school kids were infected |
|---|---|
| March 2 | The PHAS now considers the risk for infectivity in the Swedish society as moderate (increasing from low on a five-point scale; "very low", "low", "moderate", "high", "very high") |
| March 6 | The PHAS recommends to avoid non-essential travel to northern Italy. The ministry for foreign affairs in Sweden provides the same recommendation (for both northern Italy and South Korea) |
| March 10 | The PHAS notes signs of societal infectivity in Stockholm and Gothenburg metropolitan areas and increases the risk level for general societal infectivity to the highest level ("very high"). All individuals with symptoms of respiratory tract infection are encouraged to avoid social contacts that increases the risk of spread of the infection |
| March 11 | The PHAS recommends against gatherings of more than 500 individuals |
| March 11 | The WHO declares the spread of COVID-19 a pandemic. This implies no changes in the adopted Swedish policies and recommendations |
| March 13 | The PHAS declares that the COVID-19 situation in Sweden has entered a new phase. The recommendation is that all citizens with a cold or influenza-like symptoms shall remain at home |
| March 16 | The PHAS recommends individuals older than 70 to limit contact, as far as possible, with other persons. Intergenerational contacts within the family between the oldest generation and the younger generations should only take place if necessary. Employers that have the possibility to let their employees work from home should consider recommending that |
| March 17 | The PHAS recommends that all education in high schools and at Universities/University colleges takes place via distance teaching (no in-classroom teaching). The Agency thinks that individuals without symptoms shall be able to continue with their usual daily activities |
| March 19 | The PHAS notes that there are clear signs of general infectivity of COVID-19 in the Swedish society, especially in the large metropolitan areas. The Agency recommends against non-essential travel within Sweden (especially to larger cities and popular recreational places) |

| March 24 | Restrictions on cafés, restaurants and bars imposed by the the PHAS (no queues, eating and drinking only when sitting at tables) |

Source: All information in this table is collected from the press releases of the Public Health Authority in Sweden, available from: https://www.folkhalsomyndigheten.se/nyheter-och-press/nyhetsarkiv/2020/januari/nytt-coronavirus-upptackt-i-kina/

**Supplementary table S2.** Major social distancing policies adopted in different countries following the Coronavirus outbreak in 2020.

| Date | Country | Policies | Ref. |
|---|---|---|---|
| Jan. 23 | China | Lockdown of Wuhan and nearby cities in Hubei province | (5) |
| March 7 | Italy | Lock down of much of the country's north (Piedmont, Lombardy, Veneto, Emilia-Romagna) | (6) |
| March 9 | Italy | Nationwide limits on travel; all sports events and outdoor gatherings forbidden; a 6 p.m. curfew on bars | (7) |
| March 11 | Denmark | Restricts assembly of over 10 people; closed schools, restaurants, libraries, and other businesses. | (8) |
| March 12 | Norway | Closure of all schools, kindergartens and universities; cultural and sports events, gyms and businesses offering hairdressing, skincare, massage, body care and tattooing banned; buffet restaurants banned; everyone arriving in Norway from outside the Nordic to enter quarantine, regardless of whether they have symptoms or not. | (9) |
| March 14 March 16 | Spain | Nationwide quarantine imposed; only essential workers allowed to go to work; everyone else only allowed to leave home to purchase food or medication, go to hospital, walk a dog; bars, restaurants, and hotels are closed across the nation. | (8) |
| March 16 | France | Full lockdown implemented; public gatherings and walks outside banned. | (8) |

| Date | Country | Measures | Ref |
|---|---|---|---|
| March 16 (in effect on March 18) | Finland | State of emergency declared; premises of schools, educational institutions, and universities closed down; contact teaching suspended; public gatherings limited to 10 or less persons; spending unnecessary time in public places should be avoided; all public museums, theatres, other cultural venues, libraries, hobby and leisure centres, swimming pools and other sports facilities, youth centres, clubs, organisations' meeting rooms, day care services for the elderly, rehabilitative work facilities and workshops will be closed; private and third-sector operators and religious communities advised to do the same; Visits to housing services for the elderly and other at-risk groups prohibited; persons over 70 must refrain from contact with others to the extent possible (quarantine-like conditions); preparations launched for the closure of Finland's borders. | (10) |
| March 22 | Germany | Nationwide lockdown imposed; gatherings of more than two people banned; shut down of shops, churches, sports facilities, bars, and clubs. | (11) |
| March 25 | UK | Full lockdown; British public ordered to stay at home; only allowed to leave their home to do essential work, exercise, or buy food or medicine; all nonessential shops, premises, and places of worship will be closed down; weddings and baptisms banned; new rules would be enforced by the police with fines imposed on those breaking them. | (12) |

**Supplementary table S3**. Share of individuals with different mobility distances from home on January 16, 2020, and on March 26, 2020.

| Distances, m | Frequency distribution (percent) | |
| --- | --- | --- |
| | January 16, 2020 | March 26, 2020 |
| >1000 | 0.361 | 0.486 |
| 1000-4999 | 0.291 | 0.275 |
| 5000-9999 | 0.150 | 0.104 |
| 10000-19999 | 0.123 | 0.0829 |
| 20000-29000 | 0.0370 | 0.0250 |
| 30000>= | 0.0377 | 0.0266 |

**Supplementary table S4**. Variables constructed from the MIND database

| Variable | Meaning |
|---|---|
| BTWXKM | X-coordinate representing mid-point of KM2 |
| BTWYKM | Y-coordinate representing mid-point of KM2 |
| OWXKM | X-coordinate representing mid-point of KM2 of user(s) home (night-position) |
| OWYKM | Y-coordinate representing mid-point of KM2 of user(s) home (night-position) |
| ODbtwDist_mean | Average maximum distance moved away from home (per km2) |
| NofPhonesPerHour | Counting the unique number of phones that are present within km2 with a specified time-frame |

**Supplementary figures**

**Supplementary figure S1**: Map of mobile phone densities; night population, 2-3AM, on January 16, 2020.

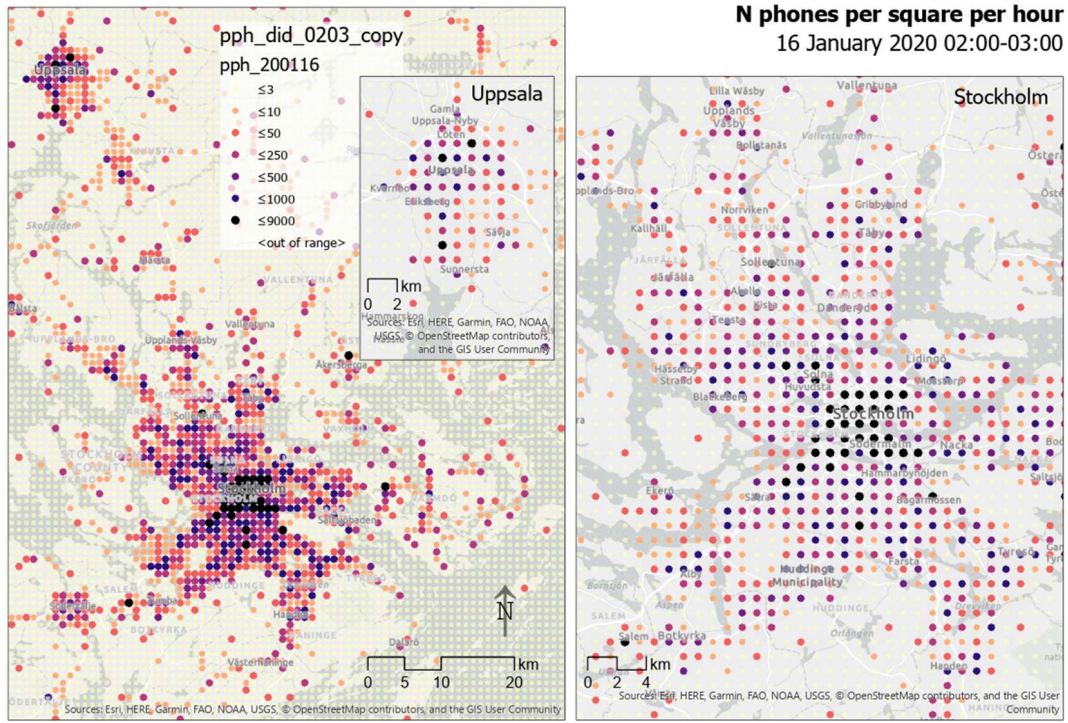

**Supplementary figure S2**: Map of mobile phone densities; day population, 10-11AM, on March 26, 2020.

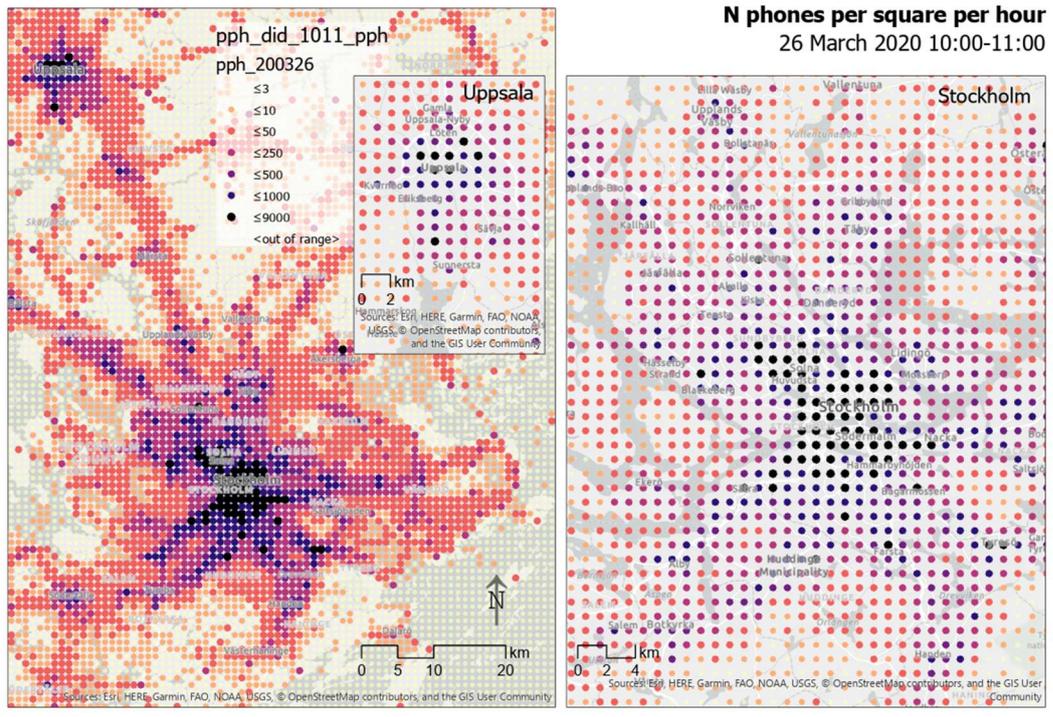

**Supplementary figure S3**. Map of land use land cover (LULC) of the study area using Corine data. The continuous urban fabric corresponds to the urban core area that includes mixed commercial and residential central areas.

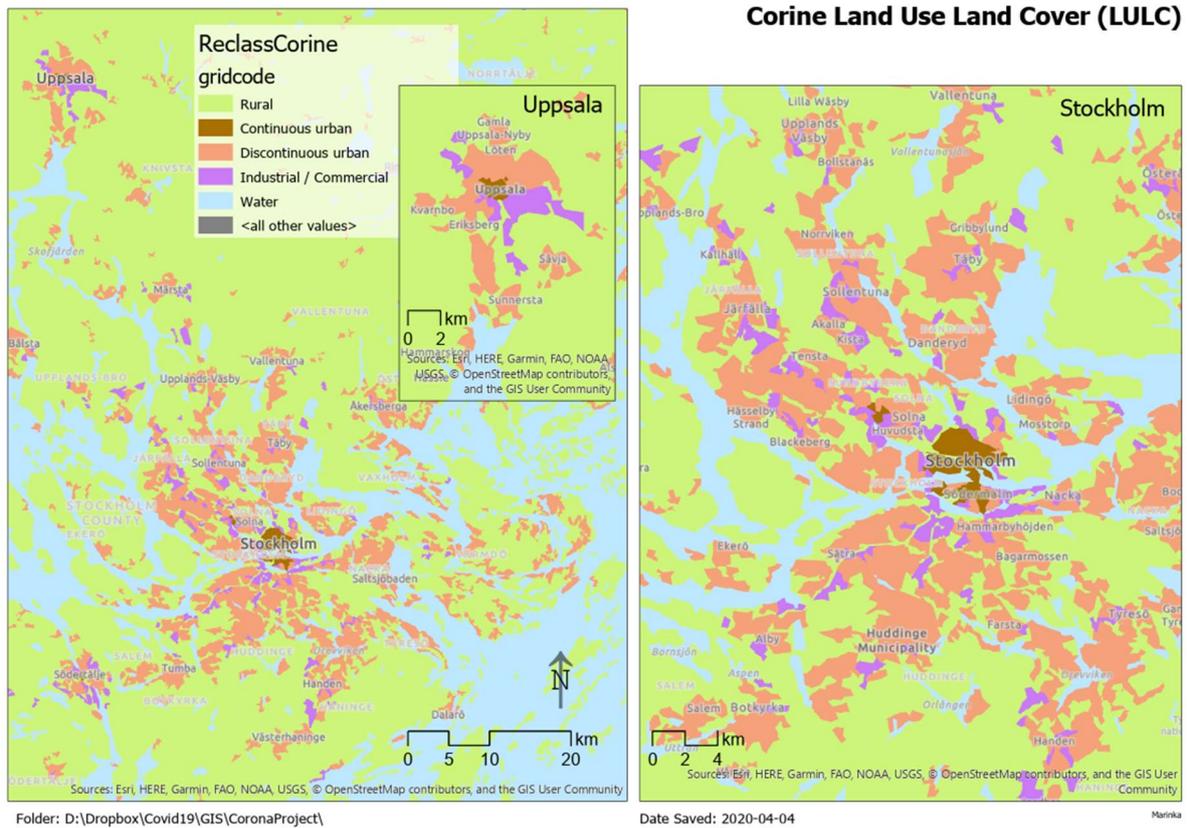

**Supplementary figure S4**: Map illustrating potential sources of bias during aggregation and allocation of geocoded data.

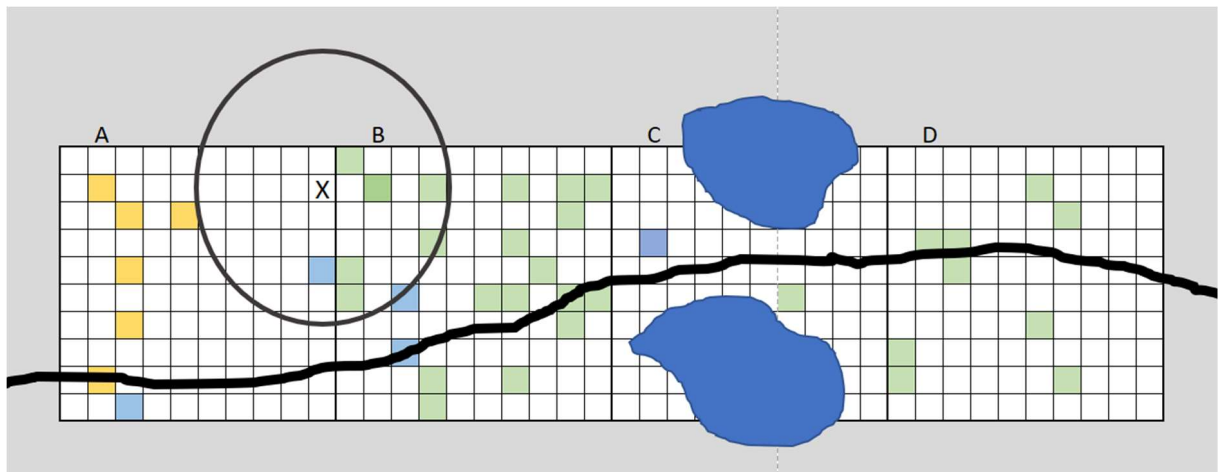